\documentclass[10pt,twoside]{article}
\usepackage{epsfig}
\usepackage{galev06}

\begin{document}



\title{Simulating Merging Galaxies:  The Infrared View}

\author{
Sukanya Chakrabarti\altaffilmark{1,2}}

\altaffiltext{1}{
Harvard-Smithsonian Center for Astrophysics, 60 Garden Street, Cambridge, MA 02138 USA, schakrabarti@cfa.harvard.edu}

\altaffiltext{2}{National Science Foundation Postdoctoral Fellow}
\begin{abstract}

We calculate multi-wavelength spectral energy distributions (SEDs) (spanning optical to millimeter wavelengths) from
simulations of major galaxy mergers using a three-dimensional radiative transfer code, which treats the absorption and scattering of radiation as well as the reemission from dust grains self-consistently.  These calculations allow us to deduce correlations from the X-rays to millimeter wavelengths.  We confirm observed correlations for Spitzer Space Telescope's IRAC bands, as well as correlations observed in the IRAS era.  We also make predictions that should be testable by future instruments.  The power of the dynamical approach afforded by calculating fluxes from the merger simulations is that we can directly correlate observed clustering in the data as seen in IRAC color-color plots with the relative amount of time the system spends in a region of color-color space.  We also present photo albums spanning the lifetime of SMGs, from its infancy in the pre-merger phase to its final stage as an elliptical galaxy, as seen in various bands.  Finally, we compare the SEDs from the simulations to recent observations of SMGs.  Our calculations, which match observed correlations both for local ULIRGs and higher redshift systems, suggest that simulations of major mergers with black hole feedback provide an excellent framework within which to understand the emission from local ULIRGs, and their high redshift
cousins, the submillimeter galaxies.
\end{abstract}
\vspace{-0.5in}
\section{Introduction}
Since the IRAS satellite detected the unexpectedly high amounts of
infrared emission generated by large numbers of dusty, infrared-bright
galaxies now commonly dubbed as LIRGs and ULIRGs (luminous infrared
galaxies, LIRGs, $L_{8~\micron-1000~\micron} \geq 10^{11} L_{\odot}$
and ultraluminous infrared galaxies, ULIRGs,
$L_{8~\micron-1000~\micron} \geq 10^{12} L_{\odot}$) (Soifer et al. 1984), 
there has been ongoing debate about the nature of the
infrared emission from these systems.  These galaxies are often
heavily obscured in the optical and radiate most of their energy at
mid and far-infrared wavelengths.  As such, much of our observational
understanding of these dusty galaxies has derived from analyses of
their infrared spectral energy distributions (SEDs).  More recently, 
submillimeter galaxies (SMGs), a relatively new class of high redshift, 
luminous ($L_{\rm IR} \ga 1\times 10^{12} L_{\odot}$), dusty galaxies,
are garnering much attention on the observational and theoretical fronts.
They were discovered in large numbers by the Submillimeter Common-User Bolometer Array (SCUBA), and generate a significant fraction of the cosmic energy output (Smail et al 1997, Blain et al 2002).   As such, SMGs are key cosmological players, believed to be responsible
for more than half the star formation at $z \sim 2$.  
These systems have been studied now in a diverse range of 
wavelengths, spanning the range from X-rays to radio wavelengths, which
suggest that these systems are very massive (Genzel et al. 2003, Tacconi et al.
2006) and host AGN (Alexander et al. 2005a). However,
many questions, such as the AGN's contribution to the luminosity and its
effect on observed correlations, as well as fundamental questions such
as whether such massive, sub-mm bright galaxies can be formed at high redshifts in
numerical simulations remain unanswered.

Observations of local ULIRGs paint a compelling picture of these galaxies
having a complex dynamical history, mediated by mergers and periods of
strong gas inflow and outflow (Downes \& Solomon 1998, Scoville et al. 2000, Surace et al. 2000, 
Rupke et al. 2005a).  This scenario is supported by simulations
demonstrating that tidal interactions during a major merger cause gas
inflows by gravitational torques (e.g. Barnes \& Hernquist 1991,
1996), leading to nuclear starbursts (e.g. Mihos \& Hernquist 1994,
1996).  

Recent merger simulations incorporating
feedback from central black holes have captured in detail their highly
dynamical evolution (Springel et al. 2005a,b) and the relation between
mergers and quasar populations in optical and x-ray observations
(Hopkins et al. 2006a,b).  Moreover, simulations like these have been
successful at reproducing massive galaxies at high redshifts (Nagamine
et al. 2005a,b).  Chakrabarti et al.
(2006a) employed these simulations of major mergers with black hole
and starburst-driven feedback to calculate the infrared emission of local ULIRGs and LIRGs
using a self-consistent fully three-dimensional radiative transfer code (Chakrabarti \& Whitney, in preparation),
which treats the absorption, reemission and scattering of radiation by dust grains, taking
as input the stellar and gas density distribution from smoothed particle hydrodynamics (SPH) simulations.
These calculations reproduced observed trends such as the well-known
warm-cold IRAS classification (de Grijp et al 1985), and demonstrated
that these trends are directly driven by feedback processes.  We review
some of the salient results from Chakrabarti et al. (2006a) and 
discuss some results on SMGs, which are presented in more detail
in Chakrabarti et al. (2006b, in preparation).   
\vspace{-0.2in}
\section{Local ULIRGs}
The warm-cold IRAS classification developed by de Grijp et al. (1985),
 wherein sources with warm ($F(25~\micron)/F(60~\micron)>0.2$) colors are
 used to identify AGN, has been used extensively since the IRAS days.  
Chakrabarti et al. (2006a) contrasted the SEDs of simulations performed 
with AGN feedback to simulations performed with starburst driven wind feedback.  
They found that the feedback processes critically determine the evolution of the
SED.  Changing the source of illumination (whether stellar or AGN) had
virtually no impact on the reprocessed far-infrared SED.  
AGN feedback is particularly effective at dispersing gas and rapidly
injecting energy into the ISM.  The observational signature of such
powerful feedback is a warm SED.  
Figure 1 shows the evolution of the $F(25~\micron)/F(60~\micron)$
colors as a function of time.  The AGN simulation is generally
warmer than the simulations with starburst feedback, and particularly
so in the most luminous phase ($t=1.2~\rm Gyr$).  The warm SED correlates with the
strong outflow phase, which suggests that feedback effects are
responsible for the increase in the high frequency emission.  
Figure 1 also shows that there is a general trend for the SEDs to
become warmer as the mass loading efficiency of the starburst winds
increases - this trend also suggests that the warm-ness of the
spectrum owes to feedback.  At late times, all the simulations lose
cold gas mass, either owing to gas dispersal by the outflow or to
new star formation - this eventually leads to the colors becoming
colder since there is not as much gas left to heat up to high
temperatures.  The exact phase of evolution when the SEDs will become
cold will depend on the mass of the galaxies - as more available gas
will fuel the warm phase for a longer time. 
\vspace{-0.2in}
\section{SMGs}
We plot in Figure 2a an IRAC color-color plot in the rest-frame.  This IRAC color-color plot was proposed by Lacy et al. (2004) to identify AGN; the AGN are the sources that fall within the dashed lines.  As Figure 2a shows, we recover the general trends of the observed IRAC color-color plot - namely, the clustering of the fluxes in the lower left hand corner of the plot (both colors being bluer), as well as a small fraction of these sources falling on the redder sequence.  We show a much smaller sample here than that shown by Lacy et al. (2004), which depicts tens of thousands of sources, while we consider a dozen simulations, at about 20 different time snapshots during the active phase on average.  Specifically, of the 16,000 objects shown in the 
Lacy plot, 2000 are likely to contain AGN, while 20 were robustly identified to be
obscured AGN; of our simulated data, 5 out of about several hundred points fall in the dashed region.  The power of the dynamical approach afforded by calculating fluxes from the simulations, is that we can unfold this color-color plot to a color vs time plot for each axis, as shown in Figures 2b and 2c.  As these plot show, the clustering that is present in the observed color-color plot, which we reproduce also in the simulated IRAC color-color plot, is naturally explained in our models by the SMG spending more of its lifetime in that region of color space.  The stars dominate in their contribution to the bolometric luminosity most of the time - which is responsible for the clustering in color-color space seen in Figure 2a.  In Chakrabarti et al. (2006b), we discuss further 
that while the presence of sources in the
AGN demarcated region in the IRAC color-color plot is a sure sign of an energetically active
AGN ($L_{\rm BH}\ga L_{\rm stars}$), the converse is not necessarily true (see also Barmby et al. 2006).  Secondly, we 
find a correlation between the rest-frame $70~ \micron$ luminosity and the hard 
X-ray luminosity which spans several orders of magnitude, as shown in Figure 3.  While
we do not currently have rest-frame $70 ~\micron$ observations of SMGs, this
prediction (the correlation, as well as the relative increase in scatter at low X-ray luminosities) 
should be testable by future instruments, such as Herschel.  

Figure 4 and Figure 5 depict the time variation of the surface brightness in the observed $3.6~\micron$
and $450~\micron$ bands of a SMG at redshift 2, from its infancy in its pre-merger phase to its final stage 
as an elliptical galaxy.  Of particular note is the increase in apparent brightness in the $3.6~\micron$ band,
which is concomitant with a decrease in brightness in the $450~\micron$ band, towards
the late phases of the merger.  We also demonstrate in Chakrabarti et al. (2006b)
that the morphology as
seen in these bands is partly a function of orbital inclination, with co-planar
mergers producing more compact remnant morphologies generally, as well as more disky morphologies in the active phase.  In Figure 6, we compare the SEDs from the simulations for a range of galaxy
masses with the observed data from Kovacs et al. (2006), which provides the
most direct observational probe of the rest-frame far-infrared of high redshift sub-mm selected galaxies
so far.  We find that our simulated galaxies span the range of observed fluxes
of SMGs and fit the data reasonably well.
\vspace{-0.2in}
\section{Conclusion}
$\bullet$ Feedback, either from the AGN or starburst driven winds, is very likely the dynamical agent that is responsible for changing the shape of the far-IR SED, in particular, the $F_{\lambda}(25~\micron)/F_{\lambda}(60~\micron)$ colors.  The presence of AGN or any source of illumination cannot be inferred from the far-IR SED alone if the source of illumination is heavily obscured such that the far-IR arises predominantly from the reprocessed emission of thermally heated dust grains.

$\bullet$ Clustering in IRAC color-color space can be naturally
explained within the context of the merger simulations studied here; clustering
in this context translates to the system spending more of its lifetime
in a given region of color-color space

$\bullet$ We predict a correlation between the rest-frame $70~\micron$ luminosity
and the hard X-ray luminosity for SMGs, which will be testable by future instruments, such 
as Herschel.

$\bullet$ Our photo albums of the lifetimes of SMGs visually illustrate the 
differential variation in surface brightness between the SCUBA and IRAC bands.
Of particular note is the increase in apparent brightness in the IRAC bands,
which is concomitant with a decrease in brightness in the SCUBA bands, towards
the late phases of the merger.  

$\bullet$ We find that simulations of major gas-rich mergers produce sub-mm bright systems the SEDs of which are in good agreement with recent
multiwavelength photometry.  

\acknowledgements{SC thanks George Rybicki for helpful discussions on scattering processes.  SC especially thanks Lars
Hernquist for many helpful discussions on merger simulations and interpretations of SMGs, and Jiasheng Huang on the observations of SMGs.  SC is supported by a National Science Foundation Fellowship.}
\vspace{-0.2in}
\begin{figure}[!hb] \begin{center}
\centerline{\psfig{file=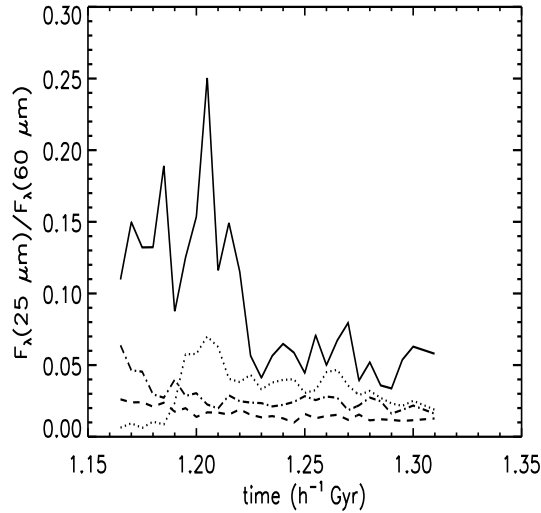,height=3.in,width=3.in}}
\end{center}
\vspace{-0.2in}
\caption{Time Evolution of $F_{25~\micron}/F_{60~\micron}$ colors for simulations
representative of local ULIRGs.  Solid line has AGN feedback, all other lines 
have varying levels of starburst feedback, with same labeling as in astro-ph/0605652}
\end{figure}
\vspace{-0.2in}
\begin{figure}[!ht] \begin{center}
\centerline{\psfig{file=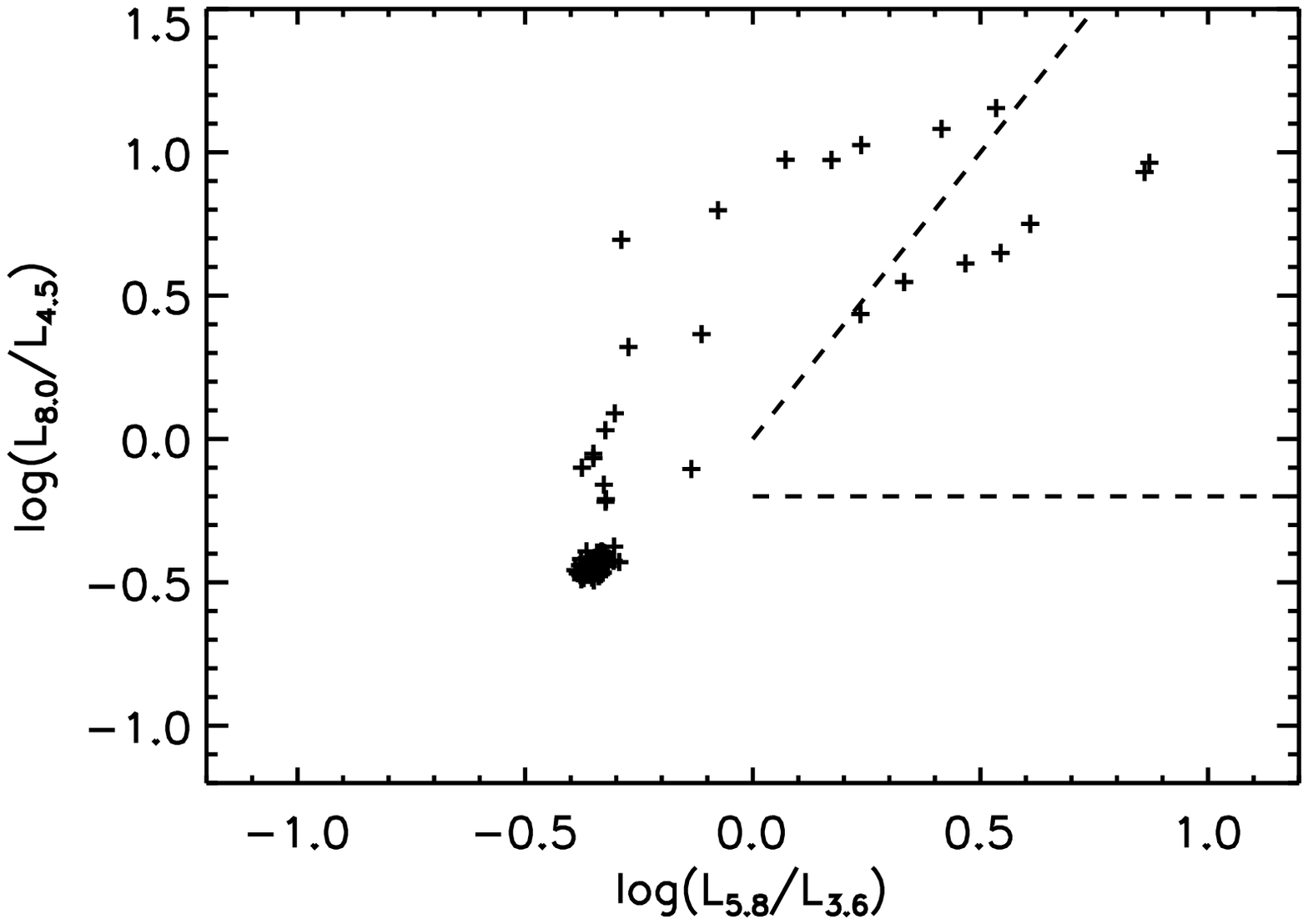,height=2.5in,width=2.5in}
\psfig{file=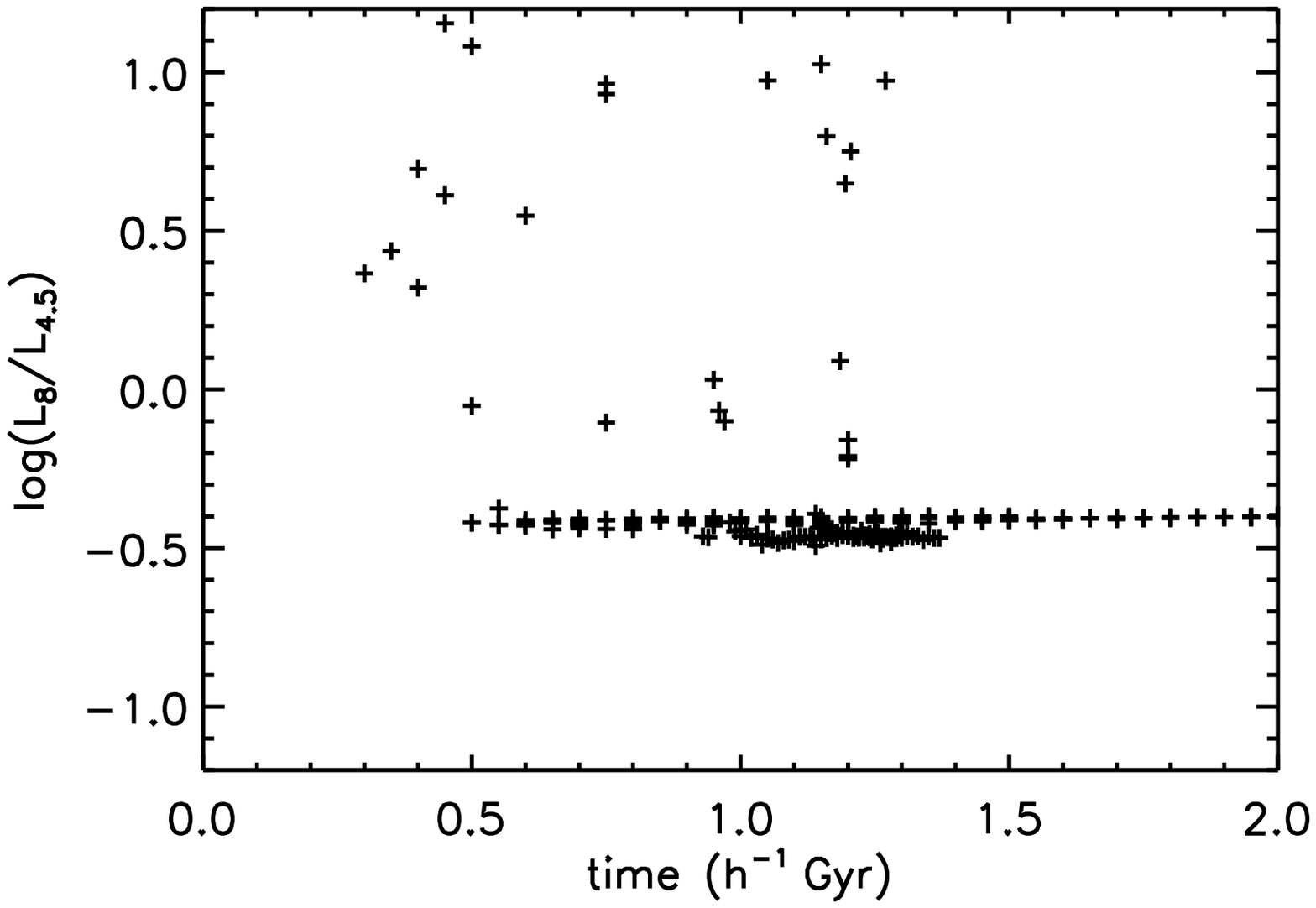,height=2.5in,width=2.5in}
\psfig{file=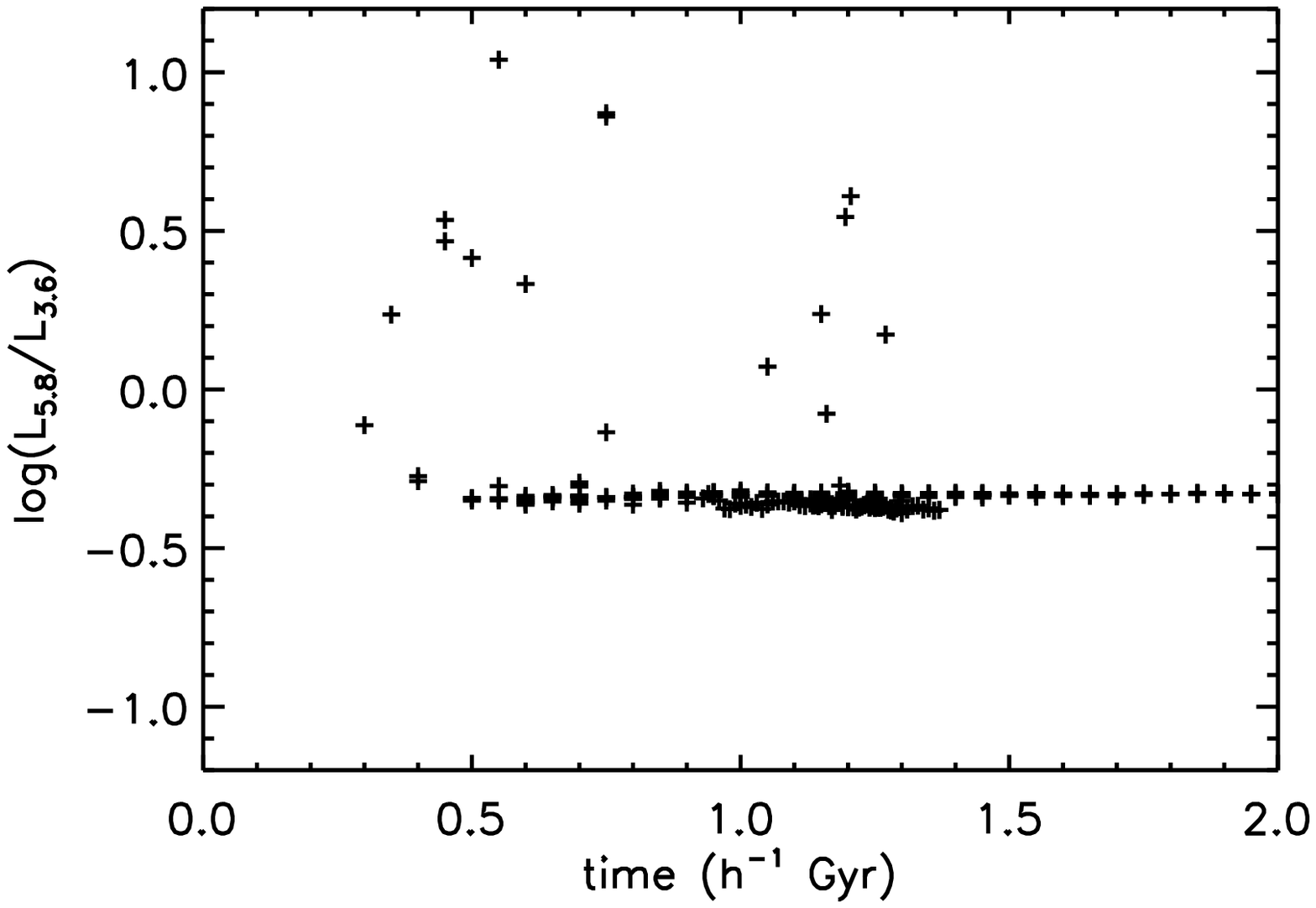,height=2.5in,width=2.5in}}
\end{center}
\caption{IRAC Color-Color plots as a function of time in the rest-frame; clustering in the lower left hand corner of the IRAC color-color plot is due to the SMG spending most of its time in that phase.}
\end{figure}
\vspace{-0.2in}
\begin{figure}[!hb] \begin{center}
\centerline{\psfig{file=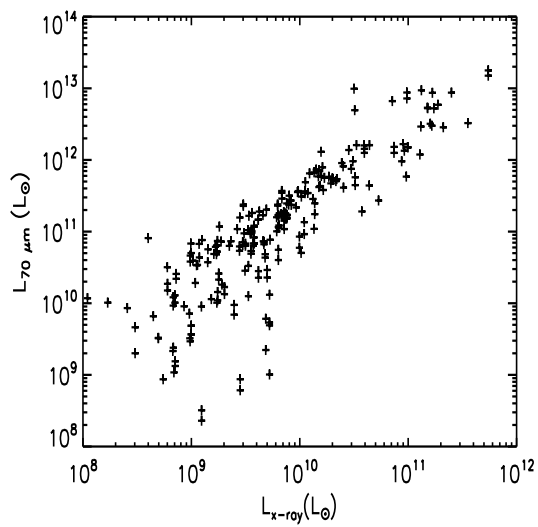,height=3.in,width=3.in}}
\end{center}
\caption{Predictions for Rest-frame $70 ~\micron$ vs. hard X-ray (2-10 keV) correlations.}
\end{figure}
\vspace{-0.2in}
\begin{figure}[p] \begin{center}
\centerline{\psfig{file=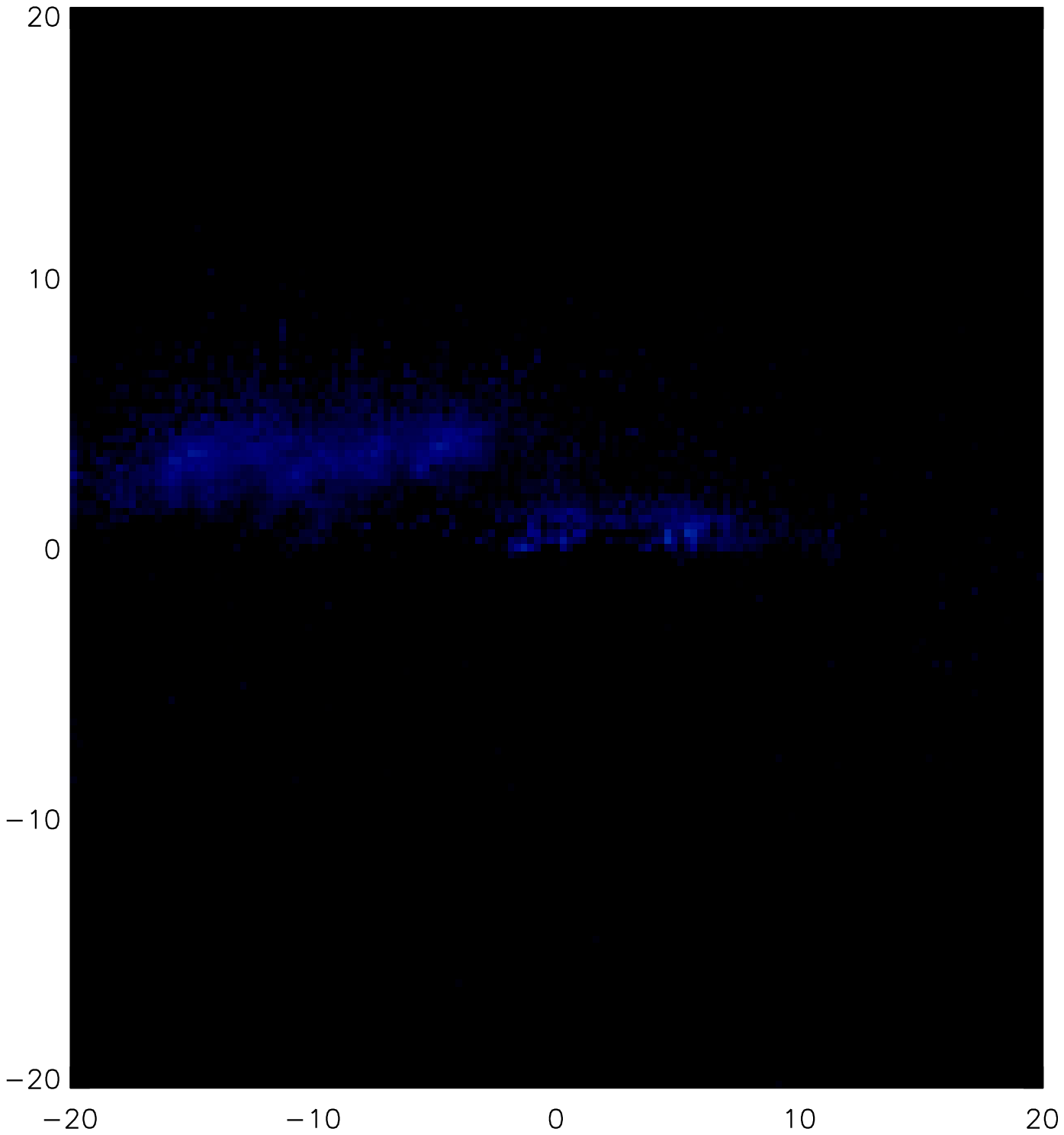,height=1.5in,width=1.5in}
\psfig{file=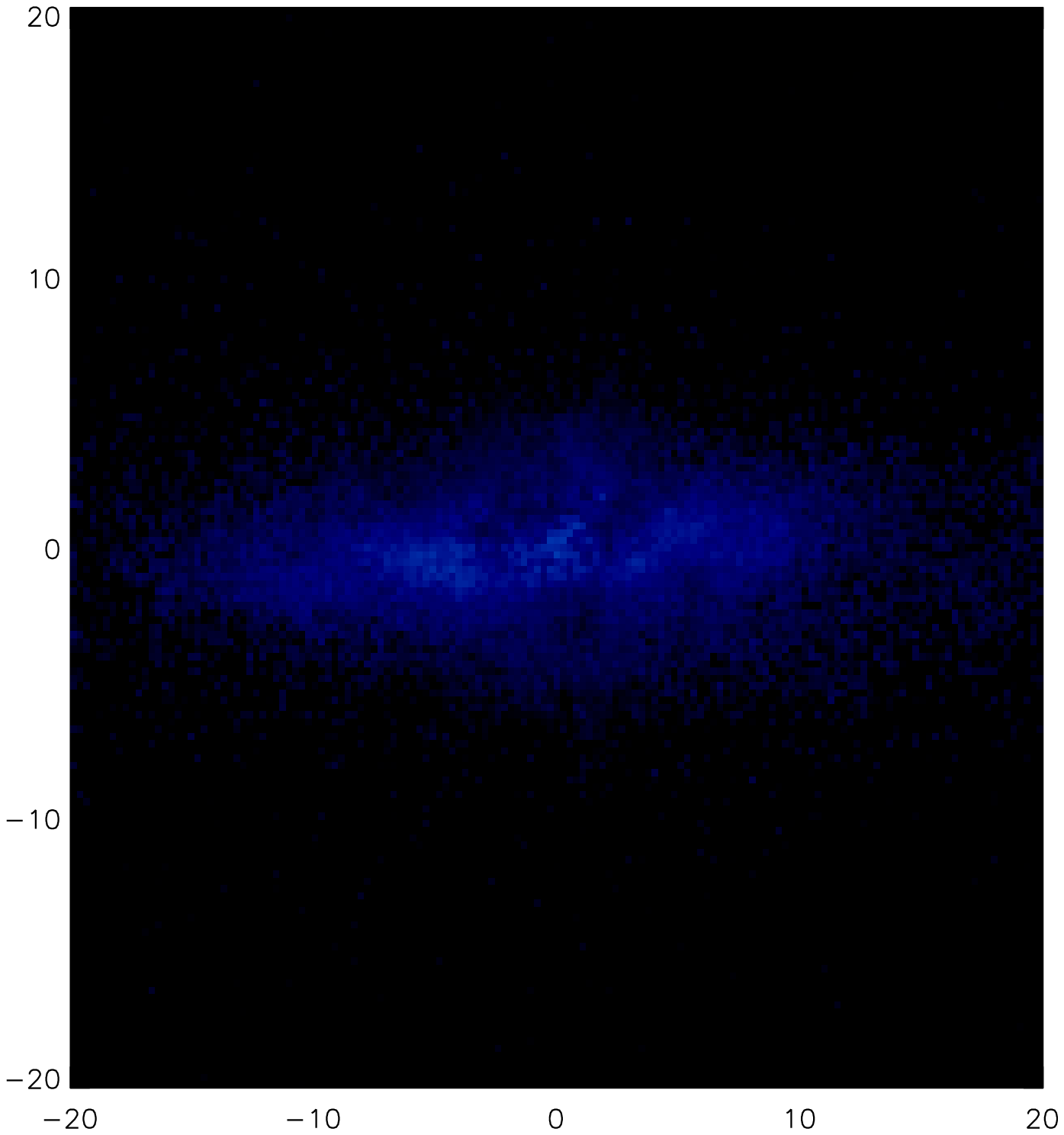,height=1.5in,width=1.5in}
{\psfig{file=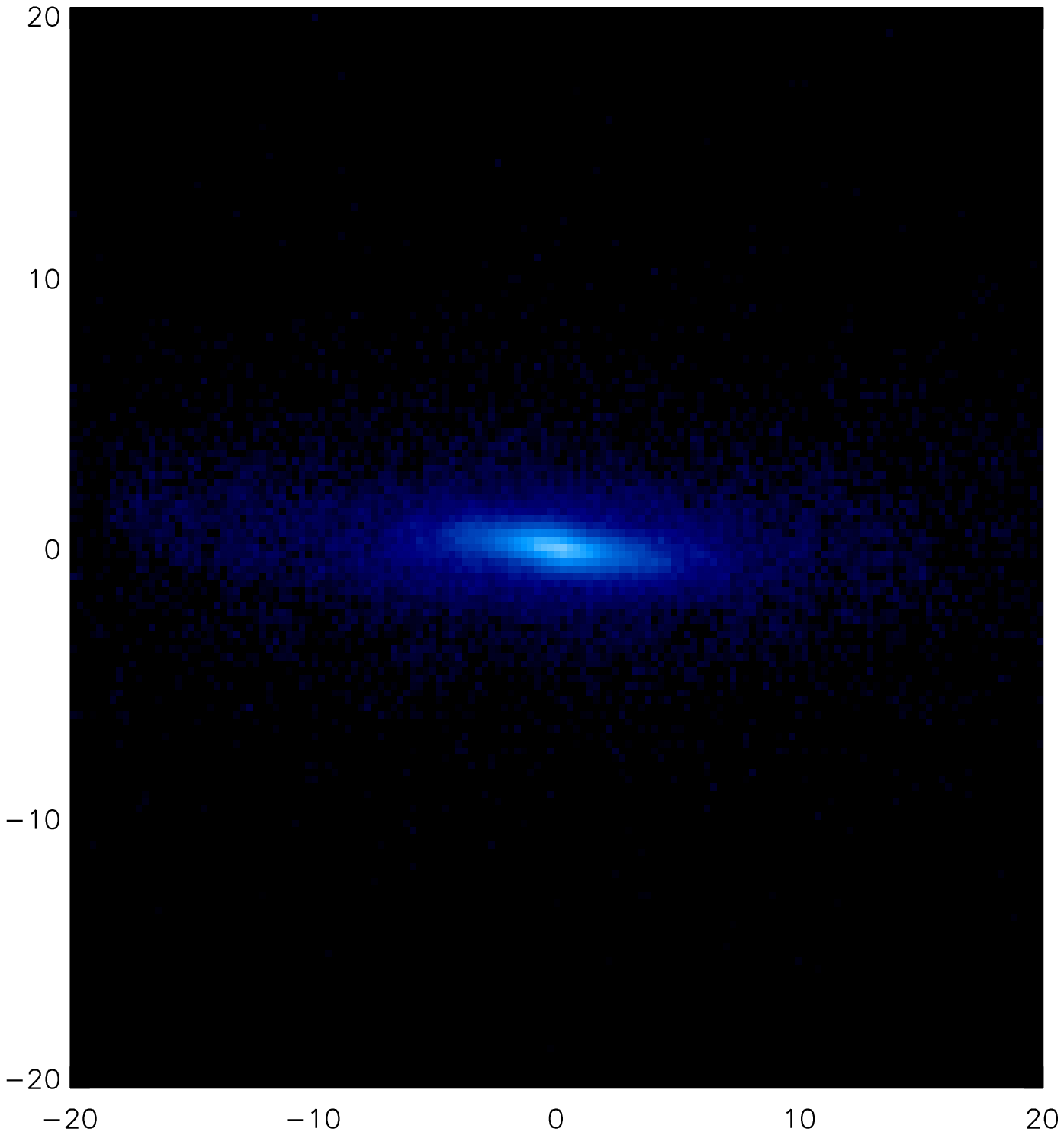,height=1.5in,width=1.5in}}}
\end{center}
\caption{Photo Album of the Lifetime of a SMG in observed $3.6~\micron$ band, (a) Pre-merger phase, (b) Close to Main Feedback Phase, (c) After Main Feedback Phase}
\end{figure}
\vspace{-0.5in}
\begin{figure}[!p] \begin{center}
\centerline{\psfig{file=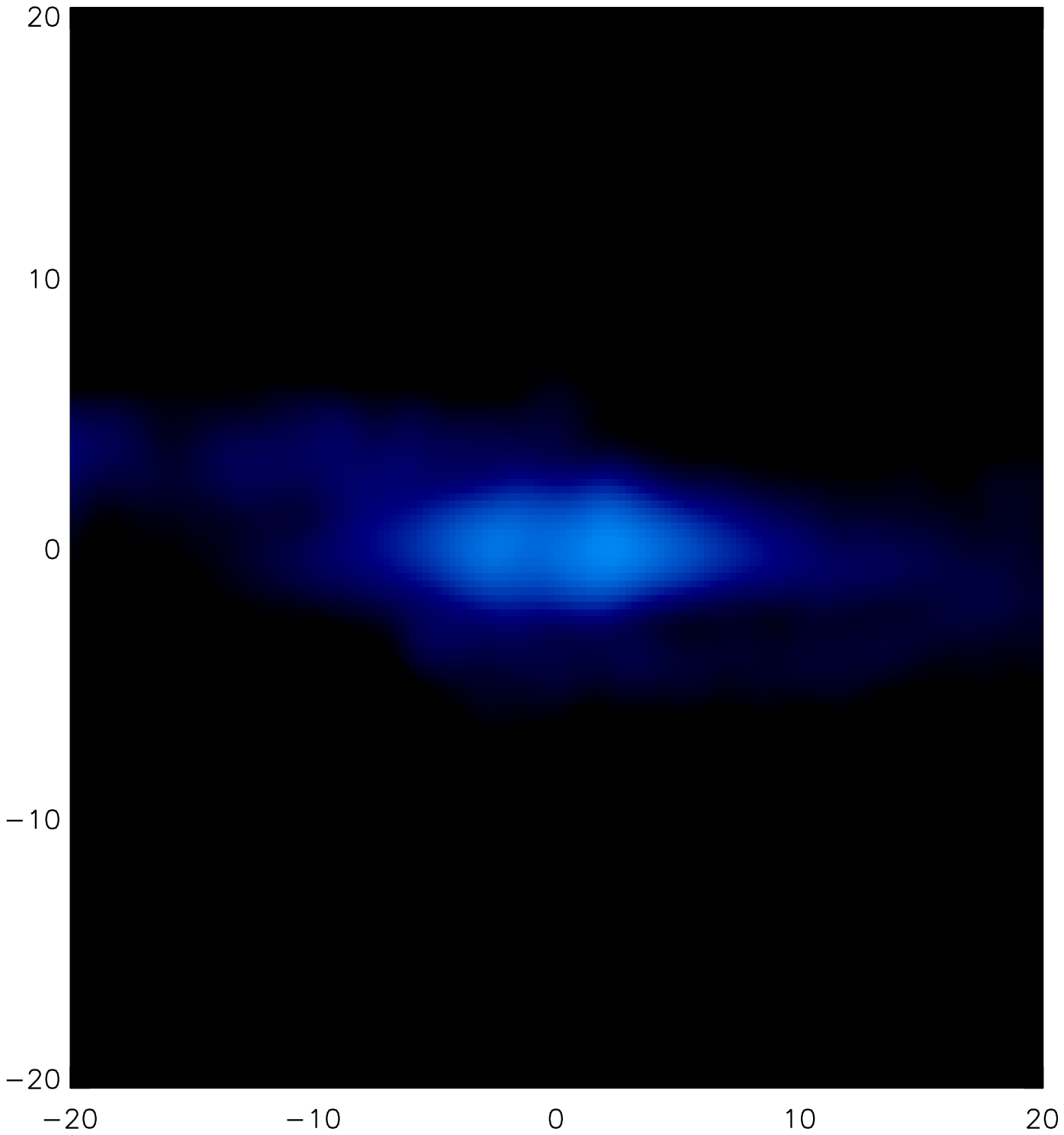,height=1.5in,width=1.5in}
\psfig{file=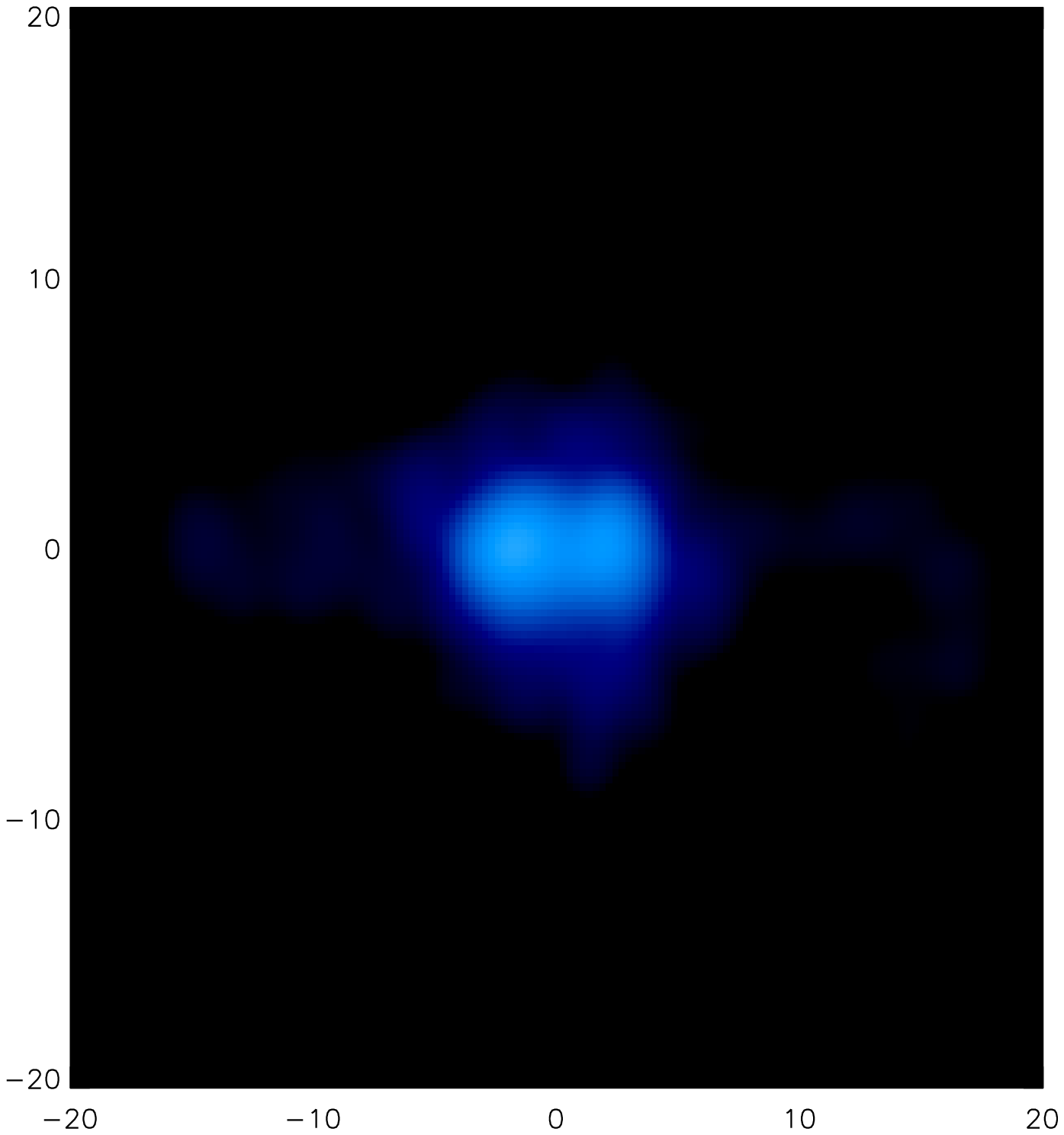,height=1.5in,width=1.5in}
{\psfig{file=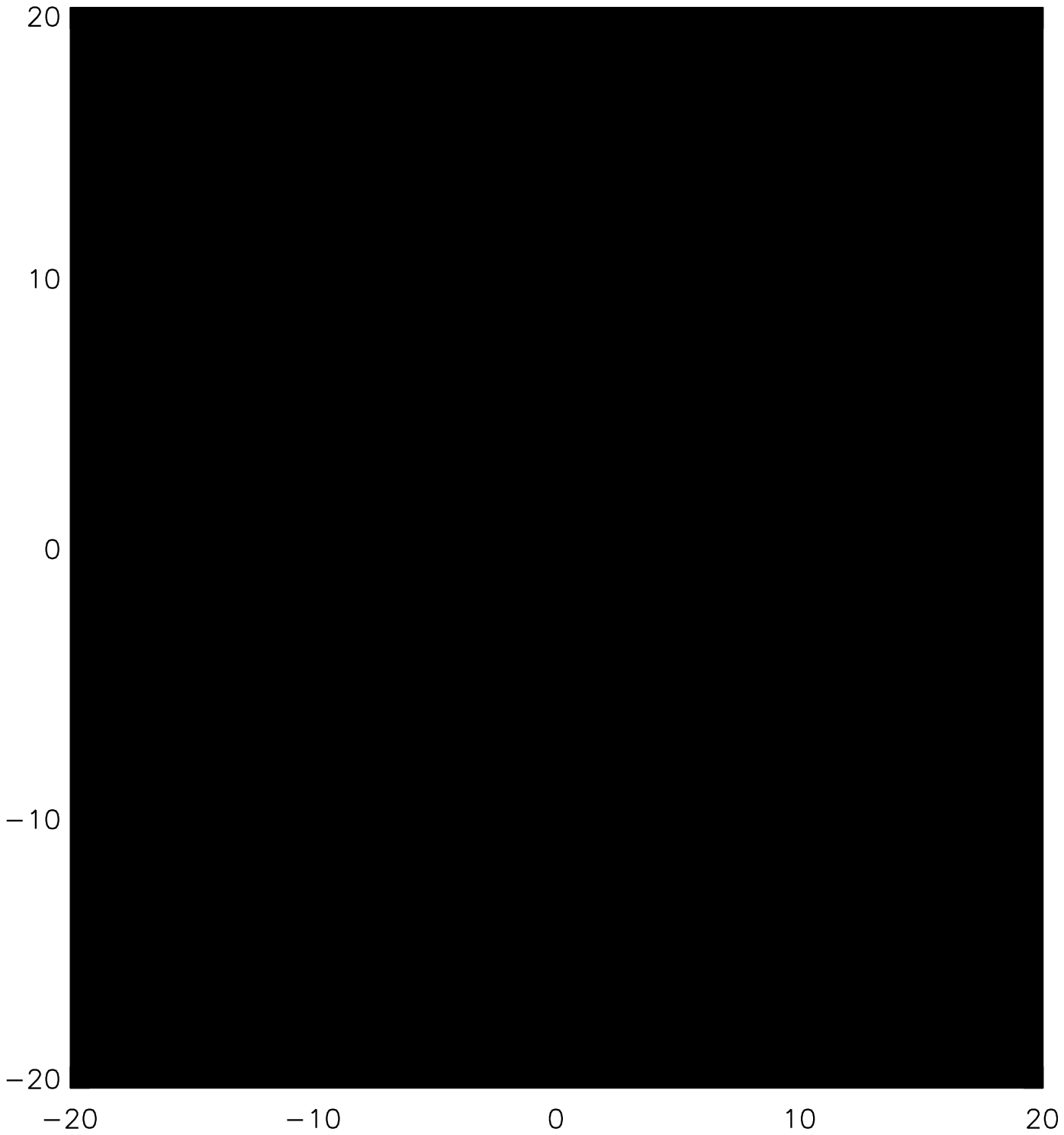,height=1.5in,width=1.5in}}}
\end{center}
\caption{Photo Album of the Lifetime of a SMG in observed $450~\micron$ band.  Times beyond the last phase are black, i.e., would show no emission.}
\end{figure}
\vspace{-0.5in}
\begin{figure}[p] \begin{center}
\centerline{\psfig{file=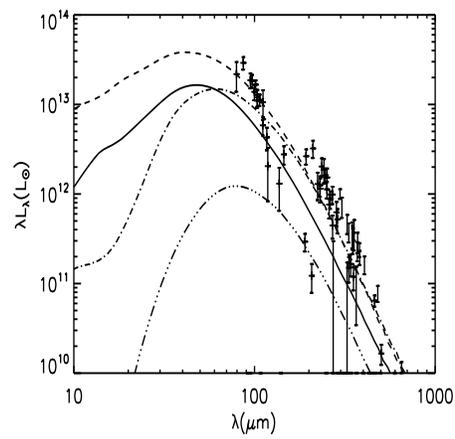,height=2.5in,width=2.5in}}
\end{center}
\caption{Comparison to SHARC-2 and SCUBA data as reported in Kovacs et al., astro-ph/0604591 (shown here in rest-frame)}
\end{figure}

\end{document}